\documentclass[11pt,a4paper]{article}
\pdfoutput=1
\usepackage{jheppub}

\usepackage{multirow, graphicx,amssymb,url,mathrsfs,amsmath}
\usepackage{wrapfig,boxedminipage,subfigure,epsfig}
\usepackage{amsxtra,amstext,latexsym,dsfont,amsfonts}
\usepackage{color}
\usepackage[dvipsnames]{xcolor}
\usepackage{float}
\usepackage{slashed}
\usepackage{calligra}
\DeclareFontShape{T1}{calligra}{m}{n}{<->s*[2.2]callig15}{}
\DeclareMathAlphabet{\mathcalligra}{T1}{calligra}{m}{n}
\usepackage{tikz} 
\usepackage[compat=1.1.0]{tikz-feynman}
\usepackage{feynmf}

\newcommand{\be}{\begin{equation}}
\newcommand{\ee}{\end{equation}}
\newcommand{\bea}{\begin{eqnarray}}
\newcommand{\eea}{\end{eqnarray}}

\title{Quantum Gravity Fluctuations in the Timelike Raychaudhuri Equation}

\author[a]{Sang-Eon Bak,}
\author[a, b]{Maulik Parikh,}
\author[c]{Sudipta Sarkar,}
\author[a]{Francesco Setti}

\affiliation[a]{Department of Physics, Arizona State University, AZ 85287, United States}
\affiliation[b]{Beyond: Center for Fundamental Concepts in Science,
Arizona State University, Tempe, Arizona 85287, USA}
\affiliation[c]{Indian Institute of Technology, Gandhinagar, Gujarat 382355, India}

\emailAdd{sbak2@asu.edu}
\emailAdd{maulik.parikh@asu.edu}
\emailAdd{sudiptas@iitgn.ac.in}
\emailAdd{fsetti@asu.edu}

\abstract{We consider a timelike geodesic congruence in the presence of perturbative quantum fluctuations of the spacetime metric. We calculate the change in the volume of a bundle of geodesics due to such fluctuations and thereby obtain a quantum-gravitationally modified timelike Raychaudhuri equation. Quantum gravity generically increases the convergence of congruences and the production of caustics.
}

\begin{document}
\maketitle

\section{Introduction}
The study of quantum fields in curved spacetime has led to the prediction of extraordinary phenomena such as the Unruh effect and Hawking radiation. In this approach, quantum matter fields are coupled to classical geometry. But an alternate situation is also of considerable interest: the study of quantum gravity coupled to classical matter. For example, the matter system could be a gravitational wave interferometer, which can be treated classically, to good approximation. More generally, the classical matter system could be a collection of massive particles that are used to probe the quantum nature of spacetime.

Now, when two quantum systems are coupled, integrating out one of the systems makes the other system behave probabilistically (i.e. stochastically) even in its classical limit. Intuitively, because the two systems are interacting, the state of the combined system generally does not factorize into a tensor product of states; tracing over the final states of the system being integrated out therefore leads to stochastic dynamics for the remaining system.

In the context of quantum gravity, a formalism has recently been developed to describe the effective dynamics of a pair of massive particles in the presence of a quantized gravitational field that has been integrated out \cite{Parikh:2020kfh,Parikh:2020fhy}; the particles experience random quantum jitters from their coupling to the underlying quantized spacetime metric \cite{Chawla:2021lop}. The geodesic separation of the particle pair no longer obeys the classical geodesic deviation equation, but instead is subject to random fluctuations. These random fluctuations, or noise, can be regarded heuristically as arising from the interaction of the particles with gravitons \cite{Parikh:2020nrd}. The statistical properties of the noise depend on the initial quantum state of the gravitational field and is explicitly computable in many cases of interest \cite{Parikh:2020fhy,Haba:2020jqs}. For certain states — in particular if the gravitational field is in a highly squeezed state — these random fluctuations may even be large enough to be potentially observable in the noisy motion of the mirrors of a gravitational wave interferometer \cite{Parikh:2020kfh,Parikh:2020fhy,Parikh:2020nrd,Kanno:2020usf,Cho_2022}.

In this paper, we extend the previous derivation from a pair of free-falling particles to a congruence of them. More specifically, we consider the evolution of a timelike congruence in the presence of a perturbatively quantized gravitational field. If spacetime were classical, such a congruence would obey the timelike Raychaudhuri equation:
\begin{equation}\label{eq_classical raychaudhuri}
\frac{d\theta}{d\tau}=-\frac{1}{3}\theta^2-\sigma_{\mu\nu}\sigma^{\mu\nu}+\omega_{\mu\nu}\omega^{\mu\nu}-R_{\mu\nu}u^\mu u^\nu\,.
\end{equation}
Here $\theta$ is the expansion, $\sigma_{\mu\nu}$ is the shear, and $\omega_{\mu\nu}$ is the vorticity of the congruence, while $R_{\mu\nu}$ is the Ricci tensor. When matter obeys the strong energy condition, Einstein's equations imply that the right-hand side in (\ref{eq_classical raychaudhuri}) becomes non-negative, causing generic congruences to converge. In a sense, the Raychaudhuri equation provides the relativistic version of the Newtonian statement that the gravitational force must be attractive. In classical general relativity, the timelike Raychaudhuri equation plays an important role in, for example, the Penrose-Hawking singularity theorem. 

The aim of this paper is to determine how the timelike Raychaudhuri equation is modified in the presence of perturbative quantum fluctuations of spacetime. Related previous work has considered spacetime fluctuations motivated by holography \cite{Verlinde:2019ade} as well as modifications to the Raychaudhuri equation arising from quantizing matter \cite{Das:2013oda,Ahmadi:2006nq,Choudhury:2021huy,Vagenas:2017fwa}. 

\section{Pairs of Particles in a Quantized Gravitational Field} \label{sec-gravity}
We are interested in the behavior of freely-falling particles in the presence of a quantized gravitational field. In this section, we briefly review the formalism of Parikh, Wilczek, and Zahariade \cite{Parikh:2020kfh,Parikh:2020fhy}; we will follow the more detailed expressions found by Cho and Hu \cite{Cho_2022}. Consider a pair of massive particles in a weak gravitational field. The action is
\begin{equation}
S[h_{\mu \nu},z] =-\frac{1}{64 \pi G_N}\int d^4x\ \partial_\mu {h}_{ij}\partial^\mu {h}^{ij} + \int dt \frac{1}{2} m \left (\delta_{ij} \dot{z}^i \dot{z}^j -\dot{{h}}_{ij} \dot{z}^i z^j \right ) \ . \label{action}
\end{equation}
Here we have expanded the Einstein-Hilbert action to second order in the metric perturbation, $h_{\mu \nu}$, where $g_{\mu \nu} = \eta_{\mu \nu} + h_{\mu \nu}$, and we are in transverse-traceless gauge. We have attached the origin of spatial coordinates to one of the particles (by working in Fermi normal coordinates), thereby reducing the problem to a system of a single particle interacting with a gravitational field. We have then expanded the action of the remaining particle to leading order in its velocity. We have also assumed that we are working in the dipole approximation, in which the metric perturbation is regarded as spatially-independent over the separation of the two particles. In (\ref{action}), $z(t)$ can be thought of as the geodesic separation of the pair of particles or, alternatively, the position of the second particle in a coordinate system affixed to the reference particle.

Next we quantize the theory. We shall, at first, regard both gravity and the particle system as quantum systems; later we will take the classical limit for the particles. Suppose now that the initial state of the gravitational field is $|\Psi\rangle$. As the field interacts with the particles, it both absorbs and emits gravitons, thereby changing its state; we therefore have to sum over final states of the field. We now would like to know the transition probability of the particle system as it goes from state $|A\rangle$ to $|B\rangle$ in time $T$:
\begin{equation}
P_{\rm \Psi}(A \to B) = \sum_{|f \rangle} | \langle f, B | \hat{U}(T) | \Psi, A \rangle |^2 \,.
\end{equation}
Here $\hat{U}$ is the unitary time-evolution operator for the combined gravity+particle system.

Since the action is quadratic, this probability can be calculated exactly. It can be regarded as a quadruple path integrals: a path integral over $z$ and a path integral over $h_{\mu \nu}$ in each of the two amplitudes making up the probability. Integrating out the gravitational field corresponds to performing the two path integrals over the metric perturbations. The result is
\begin{equation}
P_{\Psi}(A \to B) \sim \int {\cal D}z \, {\cal D}z' e^{\frac{i}{\hbar} \int_{0}^{T} dt \frac{1}{2} m (\dot{z}^2 - \dot{z}'^{2})} F_{\Psi}[z,z'] \label{probFV}
\end{equation}
where $F_\Psi[z, z']$ is known as the Feynman-Vernon influence functional \cite{Feynman:1963fq}; it encodes the entirety of the effects of the quantized gravitational field on the particle system.

The Feynman-Vernon influence functional can be exactly calculated for several important classes of states (the vacuum, coherent states, thermal states, squeezed states) of the gravitational field. For our purposes, the key observation is that the absolute value of the influence functional takes the form of the exponential of a square: $|F_\Psi[z, z']| \sim \exp(-\int \Delta^{ij}(t) K^\Psi_{ijkl}(t,t')\Delta^{kl}(t'))$, where $\Delta^{ij}(t)=z^i z^j-z'^i z'^j$ is a function of the particles' separation (for details, see \cite{Cho_2022}). Feynman and Vernon then used a trick in which they expressed the exponential as a Gaussian path integral over an auxiliary tensor field $\mathcal{N}_{ij}(t)$:
\begin{equation}\label{eq_noise term}
    e^{-\frac{m^2}{2}\int \Delta^{ij} K^{\Psi}_{ijkl}\Delta^{kl}} \sim \int \mathcal{D}\mathcal{N} \,e^{-\frac{1}{2}\int\mathcal{N}_{ij} \left(K_{\Psi}^{-1}\right)^{ijkl} \mathcal{N}_{kl}} e^{-i m \int \mathcal{N}_{ij}\Delta^{ij}}\,.
\end{equation}
Since $\mathcal{N}_{ij}(t)$ is a random function that fluctuates in a path integral, it can be thought of as noise. More precisely, it is Gaussian noise with zero mean and auto-correlation function $K^{\Psi}_{ijkl}(t,t')$. The presence of this extra path integral over $\mathcal{N}_{ij}$ means that, even after we take the classical limit by replacing the path integrals in (\ref{probFV}) with their saddle-point approximations, we are still left with the fluctuation over $\mathcal{N}_{ij}$. Thus, the classical equation of motion turns into a stochastic, rather than a deterministic, differential equation.

A more detailed calculation \cite{Parikh:2020fhy,Cho_2022,Kanno:2020usf}
shows that the influence functional also contains a phase, which has the interpretation of dissipation. This gives rise to a radiation reaction term in the equation of motion, which we will neglect.

\subsection{Quantum fluctuations in the equations of motion}
We can now take the classical limit for the dynamics of the particle separation, $z(t)$, by taking the saddle point. We find
\begin{equation}\label{eq_langevin}
\ddot{z}^{i}(t)-2 \delta^{ik}\mathcal{N}_{kl}(t)z^{l}(t)=0\,.
\end{equation}
The equation of motion has a stochastic term $\mathcal{N}_{ij}$, so it is a Langevin-type equation. The stochastic noise tensor, $\mathcal{N}_{ij}(t)$, is a matrix of random functions which has a Gaussian probability distribution with zero mean:
\begin{equation}\label{zeromean}
\langle \mathcal{N}_{ij}(t) \rangle = 0 \, .
\end{equation}
The auto-correlation function, or noise kernel, is the two-point correlation function:
\begin{equation}\label{eq_noise kernel}
    K^{\Psi}_{ijkl}(t,t')=\int \mathcal{D}\mathcal{N} P_\Psi\left[\mathcal{N}\right]\mathcal{N}_{ij}(t)\mathcal{N}_{kl}(t') \equiv \langle\mathcal{N}_{ij}(t)\mathcal{N}_{kl}(t')\rangle\,.
\end{equation}
In (\ref{zeromean}) and (\ref{eq_noise kernel}), $\langle.\rangle$ denotes the stochastic average with respect to the noise probability distribution $P_\Psi[\mathcal{N}]\sim e^{-\frac{1}{2}\int \mathcal{N}_{ij}\left(K_\Psi^{-1}\right)^{ijkl} \mathcal{N}_{kl}}$.
The statistical properties of the noise are encoded in the auto-correlation function, and depend on the quantum state of the gravitational field, $|\Psi \rangle$. Importantly, for many important classes of states, the auto-correlation function can be calculated. In particular, for the Poincar\'e-invariant vacuum state, $|0\rangle$, we have \cite{Cho_2022}
\begin{equation}\label{eq_auto}
K^{0}_{ijkl}(t,t')= -\frac{32\pi^4 \alpha^2}{15}\left(2\delta_{ij}\delta_{kl}-3(\delta_{ik}\delta{jl}+\delta_{il}\delta{jk})\right)\int^{\Lambda}_0 d\omega \omega^5 \cos{\left(\omega (t-t')\right)}\,,
\end{equation}
where $\alpha = \frac{\kappa}{2\sqrt{2}(2\pi)^3}$ with $\kappa^2= 16 \pi G_N$. For a squeezed state with uniform squeezing parameter, $r$, we have $K^{r}_{ijkl} = e^{2r} K^{0}_{ijkl}$ \cite{Parikh:2020fhy,Kanno:2020usf}. The integral in (\ref{eq_auto}) is divergent, so we need to impose an ultraviolet frequency cut-off, $\Lambda$. There are different physical motivations for the cut-off. If the massive particles correspond to the mirrors of a gravitational wave interferometer, we could choose the cut-off to correspond to the maximum frequency to which the detector is sensitive. More generally, the cut-off should be no larger than the inverse separation of the particles, as quantum fluctuations in modes that are smaller than the system will superimpose incoherently.  

We are now ready to solve the equation of motion. We take the noise to be small so that we can treat the problem perturbatively. Let $z^i(t) = z^i_0(t) + z^i_1(t) + z^i_2(t) + ...$, where the subscript on $z$ indicates the order (the number of noise factors), and the superscript indicates the spatial vector component. We find
\begin{align}
\ddot{z}^i_0(t)&=0\,,\label{eq_eom1}\\
\ddot{z}^i_1(t)&=2\delta^{ij}\mathcal{N}_{jk}(t)z^k_0(t)\,,\label{eq_eom2}\\
\ddot{z}^i_2(t)&=2\delta^{ij}\mathcal{N}_{jk}(t)z^k_1(t)\,,\label{eq_eom3}
\end{align}
Upon integrating, we have
\begin{equation}\label{eq_deviation sol}
\begin{split}
z^i(t) = z^i_0(t)&+2\delta^{ij}\int^t_0 d\tau_1 \int^{\tau_1}_0 d\tau_2 \mathcal{N}_{jk}(\tau_2)z^k_0(\tau_2)\\
&+4\delta^{ij}\delta^{kl}\int^t_0 d\tau_1 \int^{\tau_1}_0 d\tau_2\int^{\tau_2}_0 d\tau_3\int^{\tau_3}_0 d\tau_4 \mathcal{N}_{jk}(\tau_2) \mathcal{N}_{lm}(\tau_4)z^m_0(\tau_4)\,+\,\cdots.
\end{split}
\end{equation}
The first term is the solution of the zeroth-order equation (\ref{eq_eom1}), and represents the motion in the classical geometry. The second and third terms reflect the effect of quantum fluctuations of the gravitational field. The particle's motion is now stochastic, as evidenced by the presence of the random noise tensor, $\mathcal{N}$.

Taking the stochastic average of the solutions in (\ref{eq_eom1})-(\ref{eq_eom3}), we can compute the correlation functions of the fluctuation in geodesic deviations. 
By (\ref{zeromean}), we see that the one-point function of the first-order term $z^i_1(t)$ vanishes:
\begin{equation}
  \langle z^i_1(t) \rangle = 0  \, .
\end{equation}
 However, the one-point correlation function of the second-order term $z^i_2$ does not vanish:
\begin{equation}\label{eq_2pt}
\langle z^i_2(t)\rangle=4\delta^{ij}\delta^{kl}\int^t_0 d\tau_1 \int^{\tau_1}_0 d\tau_2 \int^{\tau_2}_0 d\tau_3 \int^{\tau_3}_0 d\tau_4 K^{\Psi}_{jklm}(\tau_2,\tau_4) z^m_0(\tau_4)\,.
\end{equation}
We also find a non-vanishing two-point function: 
\begin{equation}\label{eq_1pt}
\langle z^i_1(t)z^j_1(t')\rangle=4 \delta^{ik} \delta^{jl} \int^t_0 d\tau_1 \int^{\tau_1}_0 d\tau_2\int^{t'}_0 d \tau_3 \int^{\tau_3}_0 d\tau_4 K^{\Psi}_{kmln}(\tau_2,\tau_4)z^m_0(\tau_2)z^n_0(\tau_4)\,.
\end{equation}
Now that we possess expressions for correlations in the quantum fluctuations in the geodesic deviation of a pair of freely-falling massive particles, we can consider a congruence of such particles. 

\section{Congruence of Particles in a Quantized Gravitational Field}\label{sec-quantum correction}

\subsection{Volume and Raychaudhuri equation}
\begin{figure}
    \centering
    \includegraphics[width=0.6\textwidth]{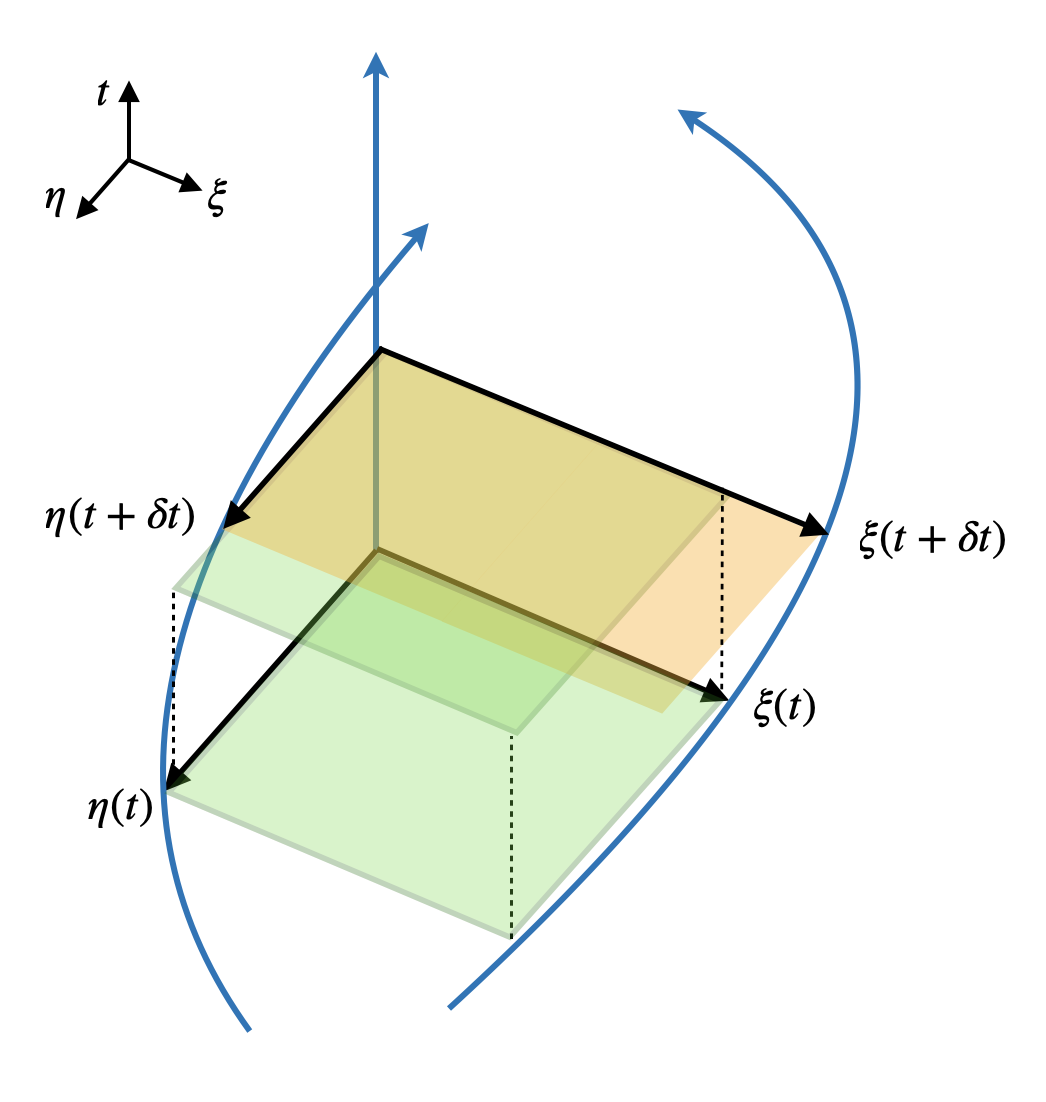}
    \caption{Schematic diagram of our setup. For simplicity, we suppress one dimension. The figure represents the evolution of the area of a congruence in two spatial dimensions. Time evolves along the vertical line. The blue lines indicate three geodesics. The black lines indicate two geodesic deviations from the reference geodesic at the origin. Note that the two geodesics lie in the $(t,\xi)$ and $(t,\eta)$ planes respectively, which have relative motion from the reference geodesic. The green bottom sheet represents the area at a certain time $t$, and the orange sheet represents the area after the time evolution by $\delta t$.}.
    \label{fig-setup}
\end{figure}

Consider now an arbitrary spacetime and focus on a small region, for which the metric can be regarded as nearly flat. Given a cut-off $\Lambda$, choose a pencil of generators of a timelike congruence whose spatial size is smaller than $\Lambda^{-1}$. We suppose that the particles are initially at rest relative to each other so that we can treat them non-relativistically. We let the pencil of generators have a cuboidal shape, for convenience. We can attach Fermi normal coordinates to one of the massive particles at one vertex of the cuboid. Then its worldline is the reference geodesic $(t,0,0,0)$, and that vertex lies at the origin of spatial coordinates. We then pick three other particles at the three vertices adjacent to the origin. Their worldlines are $(t,\xi,0,0)$, $(t,0,\eta,0)$, $(t,0,0,\zeta)$. Note that $\xi$, $\eta$, and $\zeta$ are the geodesic distances of their respective particles from the reference particle. Then the three geodesic deviation vectors define a volume in three-dimensional space (see Fig.~\ref{fig-setup}). 

To simplify the expression, we introduce the following notation:
\begin{equation}
\begin{split}
z_1^i(t)&\equiv\xi^i(t)=(\xi(t),0,0)\,,\\
z_2^i(t)&\equiv\eta^i(t)=(0,\eta(t),0)\,,\\
z_3^i(t)&\equiv\zeta^i(t)=(0,0,\zeta(t))\,,
\end{split}
\end{equation}
where $i$ runs from $1$ to $3$. Now we would like to take quantum gravitational fluctuations into account. In analogy with (\ref{eq_deviation sol}), we expand the vectors in powers of the noise:
\begin{equation}
\begin{split}
\xi(t)&=\xi_0(t)+\xi_1(t)+\xi_2(t)+\cdots\,,\\
\eta(t)&=\eta_0(t)+\eta_1(t)+\eta_2(t)+\cdots\,,\\
\zeta(t)&=\zeta_0(t)+\zeta_1(t)+\zeta_2(t)+\cdots\,.
\end{split} \label{expansion_FN}
\end{equation}
where $\xi_0(t), \eta_0(t), \zeta_0(t)$ represent the motions with constant velocities in general because they obey the equation of motion for the zeroth order (\ref{eq_eom1}). Since our particles are non-relativistic, we can disregard terms that depend both on quantum-gravity fluctuations and relativistic corrections. Therefore, we simply set the velocities to zero in the following calculations to focus on the quantum correction terms in the expansion in (\ref{expansion_FN}). At the end, we will put the velocities back in for the classical contribution. Eq. (\ref{expansion_FN}) is a series in the power of the noise factor. The dimensionless expansion parameter is $\kappa/t \sim t_p/t$, where $\kappa^2 = 16 \pi \hbar G/ c^5$ and $t_p$ is the Planck time. For any macroscopic time $t$, we have $\kappa/t \ll 1$ and hence we can neglect the higher-order terms in the expansion.

Since the equations of motion (\ref{eq_eom1}), (\ref{eq_eom2}), (\ref{eq_eom3})  are independent, the different geodesic deviations from the reference geodesic have the same solutions. Thus, we can use the same auto-correlation function when we compute the correlation functions of different geodesic deviations. We will assume that the gravitational field is in its vacuum state (the same result would hold for coherent states, while squeezed states would have an exponential enhancement in the overall factor). By using  (\ref{eq_auto}), we obtain the two-point correlation functions as follows
\begin{equation}\label{eq_two point}
\begin{split}
\langle \xi_1(t)\eta_1(t)\rangle &= -\frac{\kappa^2 \xi_0 \eta_0}{120 \pi^2 t^2} \left[(\Lambda t)^4+4(\Lambda t)^2+24+8\left((\Lambda t)^2-3\right)\cos{(\Lambda t)}-24\Lambda t \sin{(\Lambda t)}\right]\,,\\
\langle \eta_1(t)\zeta_1(t)\rangle &= -\frac{\kappa^2 \eta_0 \zeta_0}{120 \pi^2 t^2} \left[(\Lambda t)^4+4(\Lambda t)^2+24+8\left((\Lambda t)^2-3\right)\cos{(\Lambda t)}-24\Lambda t \sin{(\Lambda t)}\right]\,,\\
\langle \zeta_1(t)\xi_1(t)\rangle &= -\frac{\kappa^2 \zeta_0 \xi_0}{120 \pi^2 t^2} \left[(\Lambda t)^4+4(\Lambda t)^2+24+8\left((\Lambda t)^2-3\right)\cos{(\Lambda t)}-24\Lambda t \sin{(\Lambda t)}\right]\,,
\end{split}
\end{equation}
where the constant $\kappa^2= 16 \pi G_N$, and $\Lambda$ is the UV cutoff introduced in (\ref{eq_auto}). The two-point functions in (\ref{eq_two point}) have negative values for all time. This reflects the fact that two different geodesic deviations are negatively correlated. For example, if the first geodesic deviation $\xi^1(t)$ stretches, the second geodesic $\eta^2(t)$ shrinks in the orthogonal direction to $\xi^1(t)$. The one-point function for the higher-order terms is given by
\begin{equation}
\begin{split}
\langle \xi_2 (t)\rangle &= \frac{\kappa^2 \xi_0}{48 \pi^2 t^2} \left[-(\Lambda t)^4+12 (\Lambda t)^2+40+8\left((\Lambda t)^2-5\right)\cos{(\Lambda t)}-40\Lambda t\sin{(\Lambda t)}\right]\,,\\
\langle \eta_2 (t)\rangle &= \frac{\kappa^2 \eta_0}{48 \pi^2 t^2} \left[-(\Lambda t)^4+12 (\Lambda t)^2+40+8\left((\Lambda t)^2-5\right)\cos{(\Lambda t)}-40\Lambda t\sin{(\Lambda t)}\right]\,,\\
\langle \zeta_2 (t)\rangle &= \frac{\kappa^2 \zeta_0}{48 \pi^2 t^2} \left[-(\Lambda t)^4+12 (\Lambda t)^2+40+8\left((\Lambda t)^2-5\right)\cos{(\Lambda t)}-40\Lambda t\sin{(\Lambda t)}\right]\,.
\end{split}
\end{equation}
Now, the volume of the cuboid is merely the product of geodesic deviations:
\begin{equation}
V(t)=\xi(t) \eta(t) \zeta(t)\,. \label{voluneex}
\end{equation}
Without a quantum effect from the fluctuation of the gravitational fields, the classical volume is given by $V_c=\xi_0 \eta_0 \zeta_0$, which is time-independent. By taking the stochastic average of the volume, we find
\begin{equation}\label{eq_volume}
\begin{split}
\langle V(t) \rangle &=\xi_0 \eta_0 \zeta_0 + \xi_0 \langle \eta_1 \zeta_1 \rangle + \eta_0 \langle \zeta_1 \xi_1 \rangle + \zeta_0 \langle \xi_1 \eta_1 \rangle + \xi_0 \eta_0 \langle \zeta_2 \rangle + \eta_0 \zeta_0 \langle \xi_2 \rangle + \zeta_0 \xi_0 \langle \eta_2 \rangle + \cdots\\
&\equiv V_c(t) + V_q(t)\,.
\end{split}
\end{equation}
Here we used the fact that the terms linear in $\mathcal{N}_{ij}$ vanish. The leading contribution of quantum corrections to the volume appears in the second-order terms. If we consider the vacuum state in Minkowski space, the quantum correction to the volume is obtained as
\begin{equation}\label{eq_quantum V}
V_q(t)=\frac{\kappa^2 \xi_0 \eta_0 \zeta_0}{80\pi^2 t^2}\left[-7(\Lambda t)^4+24\left(\Lambda t\right)^2\cos{(\Lambda t)} +\cdots \right],
\end{equation}
where $\cdots$ denotes the sub leading terms in the power of $\Lambda t$. It is important to note that the contribution of quantum fluctuations to the volume is negative definite for $\Lambda t \gg 1$. This suggests that at late time when $t \gg \Lambda^{-1}$, the quantum volume may cancel the classical volume leading to the formation of caustics.

At this stage, unlike the classical case, we can define the quantum expansion associated with the geodesic congruence in different ways. Firstly, we can use the expression of the volume $V(t)$ to define the expansion as $\theta (t) = \langle \dot{V}(t) / V(t) \rangle$. Another possibility is to find the tangent $u^a$ to the geodesics and define the expansion as $\theta(t) = \langle \nabla_a u^a \rangle$. But, both of these definitions require us to apply differentiation on the fluctuating stochastic quantities. To avoid such operations, we first calculate the stochastic average of the volume and only then perform the differentiation.

Then, after taking time derivative and keeping track of the leading order correction in $V_q / V_c$, the expansion is given by
\begin{equation}\label{eq_theta}
\theta(t)=\frac{\dot{V}_c(t)}{V_c(t)}+\frac{d}{dt}\left( \frac{V_q(t)}{V_c(t)} \right) \equiv \theta_c (t) + \theta_q (t)\,,
\end{equation}
where the quantum correction to the expansion is
\begin{equation}\label{eq_quantum theta}
\theta_q(t)=-\frac{\kappa^2}{40 \pi^2 t^3}\left[7 (\Lambda t)^4+12\left(\Lambda t\right)^3\sin{(\Lambda t)} +\cdots \right]\,.
\end{equation}
In our setup, the classical expansion is zero. If the classical geodesics are initially stationary in flat spacetime, then the classical geodesics do not have the expansion. If we consider the non-zero classical expansion, it is expected that the quantum contribution to the classical expansion is very small and has the qualitatively same behavior as the quantum contribution in the zero classical expansion case. 

The expansion evolves in time as follows
\begin{equation}\label{eq_theta dot}
\dot{\theta}(t)=\frac{d}{dt}\left(\frac{\dot{V}_c(t)}{V_c(t)}\right)+\frac{d^2}{dt^2}\left(\frac{V_q(t)}{V_c(t)}\right) \equiv \dot{\theta}_c+\dot{\theta}_q\,
\end{equation}
which yields the quantum correction by
\begin{equation}\label{eq_quantum theta dot}
\dot{\theta}_q(t)=-\frac{\kappa^2}{40 \pi^2 t^4}\left[7(\Lambda t)^4+12\left(\Lambda t\right)^4 \cos{(\Lambda t)} +\cdots \right].
\end{equation}
Now $\dot{\theta}_c$ obeys the classical Raychaudhuri equation. We therefore find that the quantum-gravity-modified timelike Raychaudhuri equation is
\begin{equation}\label{eq_Raychaudhuri}
\dot{\theta}(t)=-\frac{1}{3}\theta_c^2-\sigma^2+\omega^2-R_{\mu\nu}u^{\mu}u^{\nu}-\frac{\kappa^2}{40\pi^2}\left(7\Lambda^4 + 12 \Lambda^4 \cos{(\Lambda t)} +\cdots \right)\,.
\end{equation}
Here $\kappa^2= 16 \pi G_N$, and we have restored the terms from the curvature tensor and non-zero classical expansion using the classical Raychaudhuri equation. Our equation is valid when we assume that the quantum correction is smaller than the classical volume. Therefore, we cannot use (\ref{eq_Raychaudhuri}) to analyze global properties like the formation of caustics, and instead should directly use the expression (\ref{eq_quantum V}) for the volume at time $t$.

\subsection{Caustic formation}
In our calculation, the effect of the fluctuating gravitational field is to yield a negative expansion of the geodesic congruence, which implies a kind of quantum focusing conjecture. In particular, as mentioned before, the quantum volume is always non-positive and decreases further with time. Hence, the total volume can go to zero, even when the classical volume is constant. That is, quantum gravity fluctuations enhance the production of caustics.

\begin{figure}
    \centering
    \includegraphics[width=0.5\textwidth]{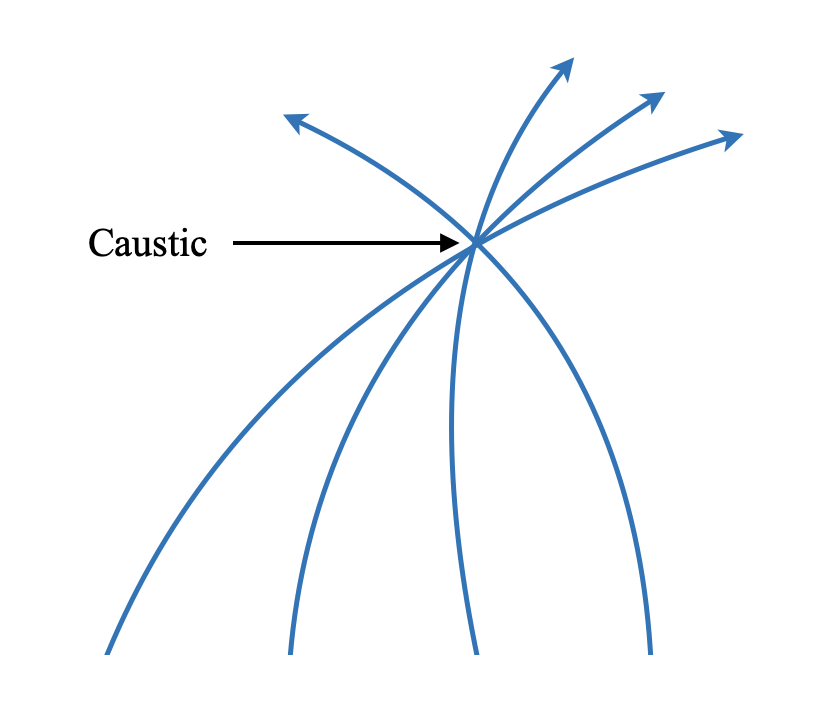}
    \caption{Due to quantum fluctuations in the spacetime metric, timelike geodesics can converge to a caustic even when the particles are initially at rest with respect to each other.}
    \label{fig-caustic}
\end{figure}

It is amusing to estimate the time-scale for this. Consider, say, a collection of massive particles in Minkowski space, which are initially at rest with respect to one another. Classically, the volume of their congruence bundle would remain constant (in the absence of their mutual gravitational attraction, which we temporarily neglect). If the initial volume is $V_0$, we have $V_c(t) = V_0$. To estimate the effects of quantum gravity, we will need to choose a cut-off, $\Lambda$. For the dipole approximation to hold, we need $\Lambda=\frac{2\pi c}{V_0^{1/3}}$. At late times, we have $\Lambda t \gg 1$, which is consistent with our earlier requirement that $\kappa/t \ll 1$. Moreover, since $\kappa \ll \Lambda^{-1}$, we can neglect the higher-order terms in $\kappa$. Then, using equation (\ref{eq_quantum V}) with $V_q \equiv - V_0$, we find the time until a caustic is formed to be
\begin{equation}\label{eq_caustic}
t_c=\sqrt{\frac{5 \,c}{112\pi^3 G_N \hbar}} V_0^{2/3} \sim t_p \left ( \frac{\text{Area}}{l_p^2} \right )\,,
\end{equation}
where the last expression indicates the scaling behavior of the gravitational collapse time with the Planck time $t_p$ and the Planck length $l_p$. Although the quantum-gravitational collapse time scales with the area (if we consider the initial volume in a symmetric way, i.e., $V_0^{2/3}=\text{Area}$), the result seems unlikely to have a holographic interpretation, as our calculation uses only perturbative quantum gravity; indeed, the collapse time does not scale as a co-dimension two quantity in general spacetime dimensions. Inserting some numbers into (\ref{eq_caustic}), we find for example that a cube of particles a nanometer across would collapse in about three months through quantum gravity fluctuations. By comparison, the classical gravitational collapse time for particles of mass $m$ goes as $\sim \sqrt{\frac{V_0}{2G_Nm}}$. Note that our estimate assumes that the quantum state of the gravitational field is the vacuum state. If instead the field is in a squeezed state with squeezing parameter $r$, the time to collapse decreases by a factor of $e^r$. It would be interesting to determine whether there exists observational data along these lines that might constrain $r$ \cite{Hertzberg:2021rbl,Guerreiro:2021qgk}.

\section{Conclusion} \label{sec-disc}
In this paper we used the quantum-gravity-induced stochastic fluctuations of pairs of timelike geodesics to analyze the behavior of a congruence of such geodesics. We found that quantum fluctuations generically cause the congruence volume to decrease relative to its classical evolution. In particular, the worldines of particles that are at rest in Minkowski space will eventually converge to a caustic. Finally, working in the limit that the quantum contribution to the volume is small compared with the classical volume, we were able to determine the quantum-gravity modification to the timelike Raychaudhuri equation.

There are several promising directions for future work. The Raychaudhuri equation determines the evolution of the congruence expansion, $\theta$, but there are also equations for the evolution of the shear and vorticity. Symmetry arguments suggest that there should be no quantum-gravitational corrections to the vorticity equation. However, there should be quantum-gravitational contributions to the evolution equation for the shear, and it would be interesting to determine that. An important direction would be to consider the quantum-gravity modified evolution equations for null congruences. It would be interesting to compare the results of this approach, which relies on perturbative quantum gravity, with equations obtained using holographic arguments about generalized entropy, such as the quantum focusing conjecture and the quantum singularity theorem \cite{Bousso:2015mna, Bousso:2022tdb}.\\

\noindent
{\bf Acknowledgments}\\
We thank George Zahariade for helpful discussions. MP is supported in part by Heising-Simons Foundation grant 2021-2818, Department of Energy grant DE-SC0019470 and Government of India DST VAJRA Faculty Scheme
VJR/2017/000117. The research of S.S. is supported by the Department of Science and Technology, Government of India under the SERB
CRG Grant CRG/2020/004562.


\bibliographystyle{JHEP}

\end{document}